\documentclass{ecai} 

\usepackage[nolist, acronym]{glossaries}
\usepackage[hidelinks]{hyperref}
\usepackage[numbers]{natbib} 
\usepackage{latexsym}
\usepackage{amssymb}
\usepackage{amsmath}
\usepackage{amsthm}
\usepackage{booktabs}
\usepackage{enumitem}
\usepackage{graphicx}
\usepackage{color}
\usepackage{cleveref}

\newcommand{\BibTeX}{B\kern-.05em{\sc i\kern-.025em b}\kern-.08em\TeX}

\makeglossaries
\newacronym{ai}{AI}{Artificial Intelligence}
\newacronym{aoi}{AOI}{Area of Interest}
\newacronym{bilstm}{BiLSTM}{Bidirectional Long Short-Term Memory}
\newacronym{cot}{CoT}{Chain of Thought}
\newacronym{ct}{CT}{Conversation Tree}
\newacronym{cnn}{CNN}{Convolutional Neural Network}
\newacronym{cv}{CV}{Computer Vision}

\newacronym{drl}{DRL}{Deep Reinforcement Learning}
\newacronym{ddpg}{DDPG}{Deep Deterministic Policy Gradient}
\newacronym{dl}{DL}{Deep Learning}
\newacronym{dnn}{DNN}{Deep Neural Networks}
\newacronym{dqn}{DQN}{Deep Q-learning}
\newacronym{dpo}{DPO}{Direct Preference Optimization}
\newacronym{ddqn}{DDQN}{Double Q-learning}
\newacronym{et}{ET}{Eye-tracking}
\newacronym{eeg}{EEG}{Electroencephalography}
\newacronym{ft}{FT}{Fine-Tuning}
\newacronym{gae}{GAE}{Generalized Advantage Estimate}
\newacronym{gnn}{GNN}{Graph Neural Networks}
\newacronym{gqa}{GQA}{Grouped Query Attention}
\newacronym{hrlaif}{HRLAIF}{Hybrid Reinforcement Learning from AI Feedback}
\newacronym{infovis}{InfoVis}{Information visualization}
\newacronym{il}{IL}{Imitation Learning}
\newacronym{ipopt}{IPOPT}{Interior Point Optimizer}
\newacronym{kl}{KL}{Kullback–Leibler}
\newacronym{lm}{LM}{Language Model}
\newacronym{lms}{LMs}{Language Models}
\newacronym{llm}{LLM}{Large Language Model}
\newacronym{llms}{LLMs}{Large Language Models}
\newacronym{lstm}{LSTM}{Long short-term memory}
\newacronym{lora}{LoRA}{Low-Rank Adaptation}
\newacronym{lr}{LR}{Learning Rate}
\newacronym{mwl}{MWL}{mental workload}
\newacronym{mdp}{MDP}{Markov Decision Process}
\newacronym{mha}{MHA}{multi-head attention}
\newacronym{ml}{ML}{Machine Learning}
\newacronym{mllm}{MLLM}{Multimodal Large Language Model}
\newacronym{mllms}{MLLMs}{Multimodal Large Language Models}
\newacronym{mlp}{MLP}{Multilayer Perceptron}
\newacronym{mpnn}{MPNN}{Message Passing Neural Networks}
\newacronym{mtl}{MTL}{Multi-task Learning}
\newacronym{ner}{NER}{Named Entity Recognition}
\newacronym{nlp}{NLP}{Natural Language Processing}
\newacronym{nn}{NN}{Neural Networks}
\newacronym{ocr}{OCR}{Optical Character Recognition}
\newacronym{orms}{ORMs}{Outcome-supervised reward models}
\newacronym{p3o}{P3O}{Pairwise Proximal Policy Optimization}
\newacronym{peft}{PEFT}{Parameter-Efficient Fine-Tuning}
\newacronym{pi}{PI}{Policy Iteration}
\newacronym{ppo}{PPO}{Proximal Policy Optimization}
\newacronym{prms}{PRMs}{Process Based Reward Models}
\newacronym{qa}{QA}{Question Answering}
\newacronym{raft}{RAFT}{Reward rAnked FineTuning}
\newacronym{rag}{RAG}{Retrieval-Augmented Generation}
\newacronym{rnn}{RNN}{Recurrent Neural Network}
\newacronym{rnns}{RNNs}{Recurrent Neural Networks}
\newacronym{rl}{RL}{Reinforcement Learning}
\newacronym{rlaif}{RLAIF}{Reinforcement Learning from AI Feedback}
\newacronym{rlcd}{RLCD}{Reinforcement Learning from Contrastive Distillation}
\newacronym{rlhf}{RLHF}{Reinforcement Learning from Human Feedback}
\newacronym{rm}{RM}{Reward Model}
\newacronym{rms}{RMs}{Reward Models}
\newacronym{rso}{RSO}{Statistical Rejection Sampling Optimization}
\newacronym{rrhf}{RRHF}{Rank Responses to align Human Feedback}

\newacronym{sota}{SOTA}{State of the Art}
\newacronym{sft}{SFT}{Supervised Fine-Tuning}
\newacronym{smoe}{SMoE}{Sparse Mixture of Experts}
\newacronym{slic}{SLiC}{Sequence Likelihood Calibration}
\newacronym{slcihf}{SLiC-HF}{Sequence Likelihood Calibration with Human Feedback}
\newacronym{sorms}{SORMs}{Stepwise ORMs}
\newacronym{swa}{SWA}{sliding window attention}
\newacronym{trpo}{TRPO}{Trust Region Policy Optimization}
\newacronym{trt}{TRT}{total reading time}

\newacronym{vi}{VI}{Value Iteration}
\newacronym{vqa}{VQA}{Visual Question Answering}

\begin{document}


\begin{frontmatter}


\paperid{123} 


\title{Context-aware Adaptive Visualizations for Critical Decision Making}


\author[A]{\fnms{Angela}~\snm{Lopez-Cardona}\thanks{Corresponding Author. Email: angela.lopezcardona@telefonica.com, nuwan.attygalle@uclouvain.be}\footnote{Equal contribution.}}
\author[A]
{\fnms{Mireia}~\snm{Masias Bruns}\footnotemark} 
\author[B]
{\fnms{Nuwan}~\snm{T. Attygalle}\footnotemark} 
\author[A]
{\fnms{Sebastian}~\snm{Idesis}}
\author[A]
{\fnms{Matteo}~\snm{Salvatori}}
\author[C]
{\fnms{Konstantinos}~\snm{Raftopoulos}} 
\author[C]
{\fnms{Konstantinos}~\snm{Oikonomou}} 
\author[D]
{\fnms{Saravanakumar}~\snm{Duraisamy}}
\author[D]
{\fnms{Parvin}~\snm{Emami}}

\author[B]
{\fnms{Nacera}~\snm{Latreche}}
\author[B]
{\fnms{Alaa Eddine Anis}~\snm{Sahraoui}}

\author[C]
{\fnms{Michalis}~\snm{Vakalellis}} 

\author[B]
{\fnms{Jean}~\snm{Vanderdonckt}}
\author[A]
{\fnms{Ioannis}~\snm{Arapakis}} 
\author[D]
{\fnms{Luis A.}~\snm{Leiva}}

\address[A]{Telef\'{o}nica Scientific Research, Spain}
\address[B]{Universit\'{e} catholique de Louvain, Belgium}
\address[C]{AEGIS IT Research GmbH, Germany}
\address[D]{University of Luxembourg, Luxembourg}


\begin{abstract}
Effective decision-making often relies on timely insights from complex visual data. While Information Visualization (InfoVis) dashboards can support this process, they rarely adapt to users' cognitive state, and less so in real time. We present \textsc{Symbiotik}, an intelligent, context-aware adaptive visualization system that leverages neurophysiological signals to estimate mental workload (MWL) and dynamically adapt visual dashboards using reinforcement learning (RL). Through a user study with 120 participants and three visualization types, we demonstrate that our approach improves task performance and engagement. \textsc{Symbiotik} offers a scalable, real-time adaptation architecture, and a validated methodology for neuroadaptive user interfaces.
\end{abstract}

\end{frontmatter}

\noindent \textbf{© 2025 IOS Press. This is the author's version of this work, provided for personal use only. Redistribution is prohibited. The definitive version is published in the Proceedings of the \emph{European Conference on Artificial Intelligence (ECAI 2025)}, DEMO Track, ISBN 978-1-64368-543-2, \url{https://doi.org/10.3233/FAIA251433}.}


\section{Introduction}
\label{sec:introduction}

In today's information-saturated environment, users must decide what to process, filter, and prioritize. Although using visualizations to address this challenge has been widely discussed~\cite{Falschlunger2016}, \gls{infovis} systems have yet to reach their full potential to support critical decision-making. These systems leverage visual elements—charts, graphs, networks, and maps—to facilitate intuitive insights by engaging perceptual mechanisms rather than requiring laborious data comparisons~\cite{wang_salchartqa_2024}. The effectiveness of \gls{infovis} in responding to critical events hinges on the visual representation of data and the speed at which actionable insights can be derived. We argue that physiological computing holds significant promise for innovation in \gls{infovis} by incorporating human body signals, which can lead to more precise user interface adaptations through closed-loop implicit monitoring~\cite{Fairclough22_neuroadaptive}. This is supported by evidence that cognitive processes are affected by user interface aesthetics, detectable through physiological signals~\cite{hirshfield_this_2011, peck_using_2013, lukanov_using_2016, Haddad24}.

We argue that future solutions should consider a user-centric approach, premised on context awareness and emotion-sensing capabilities. In this regard, we propose an innovative adaptive framework inspired by symbiosis principles, where humans and machines cooperate and co-evolve to support decision-making processes~\cite{leiva_context-aware_2023}. To this end, we envision a bidirectional channel of communication between AI and human users, which taps directly into human cognitive resources to facilitate an effortless and natural dialogue.

Recent literature highlights the growing need for adaptive visualization systems that consider individual differences in cognitive traits and mental states. According to~\citet{yanez_state_2025}, each user experiences visualizations through their own lens, shaped by factors such as past experiences, personality, and cognitive abilities--all of which can significantly influence attention, speed, and task accuracy. Furthermore,~\citet{chiossi_adapting_2022} emphasizes the importance of multi-modal adaptive systems capable of capturing the full context of the user.

Prior studies have explored physiological signals to support attention modeling and adaptive interfaces. \citet{salous_smarthelm_2022} developed \textit{SmartHelm}, a smart helmet for cargo cyclists that uses multimodal signals and LSTM models to detect attention lapses. \citet{toreini_designing_2022} proposed attentive dashboards that rely solely on real-time \gls{et} to deliver individualized visual attention feedback, improving attentional resource management during data exploration. ~\citet{karran_toward_2019} used EEG-based neurofeedback in a business context, applying an engagement index and a calibrated classifier to trigger adaptive interface changes, which led to better sustained attention and reduced error. More recently,~\citet{gehrke_neuroadaptive_2025} demonstrated a neuroadaptive haptics system that uses \gls{rl} to adapt haptic feedback in XR environments based on either explicit user ratings or EEG-based neural decoders.

Among the methodologies aimed at measuring brain activity and cognitive processes in the field of neuroimaging, \gls{eeg} stands out for its ability to capture electrical activity from the scalp with high temporal resolution, making it suitable for studying real-time neural dynamics. \gls{eeg} is a non-invasive and portable method, offering practical advantages over other techniques. By leveraging \gls{eeg} as a neurophysiological signal for adaptation, our system contributes to this vision--enabling more effective, personalized, and cognitively aligned data exploration experiences. 

In contrast to previous work—none of which has applied \gls{eeg}-based \gls{mwl} estimation to \gls{infovis}—our approach uses this signal to drive the real-time adaptation. We continuously track cognitive load and use RL agents to adapt the dashboard in real time. This enables more granular, workload-sensitive adaptations, offering a novel contribution to interface personalization. To this end, we first studied which types of adaptations in different kinds of \gls{infovis} dashboards result in enhanced performance and quicker response times in critical situations. We then trained \gls{rl} agents with this data and implemented a neuro-feedback loop with context-aware \gls{rl} agents during inference.

Based on an investigation case study involving different types of visualizations (\Cref{sec:case_study}), the proposed framework provides \gls{rl} agents (\Cref{subsec:rl_module}) with environmental awareness through an embedding space of neurophysiological descriptors (\Cref{subsec:cdr_module}), captured using commodity sensors (\Cref{subsec:physiological_sensing_module}). These descriptors drive adaptations of the dashboard interface (\Cref{subsec:adaptations_engine_module}) within a closed-loop, implicit monitoring framework (\Cref{sec:system}).

\section{Use Case}
\label{sec:case_study}

The use case we address involves the investigation of a criminal network by a Law Enforcement Agency (LEA). 
We created an adaptive Crime Investigation Dashboard (CID), built upon AEGIS's Advanced Visualization Toolkit (AVT). The formulation of relationships across individuals, the underlying factors determining these connections, and the emergence of cohesive groups (i.e. cliques) with dense links are analysed to uncover hidden patterns and understand behaviours and relations within the network.

To this end, information foraging of data involving interactions/communications within a specific period is investigated. Call Data Records (CDRs) are acquired from relevant telecommunications operators after identifying telephone numbers associated with the group i.e. person(s) of interest of specific phone numbers and devices. This is achieved by analyzing the available historical information extracted from files across multiple modalities, all grounded in the CDR-based interaction patterns previously described.

\begin{figure}[!htbp]
    \centering
    \includegraphics[width=\linewidth]{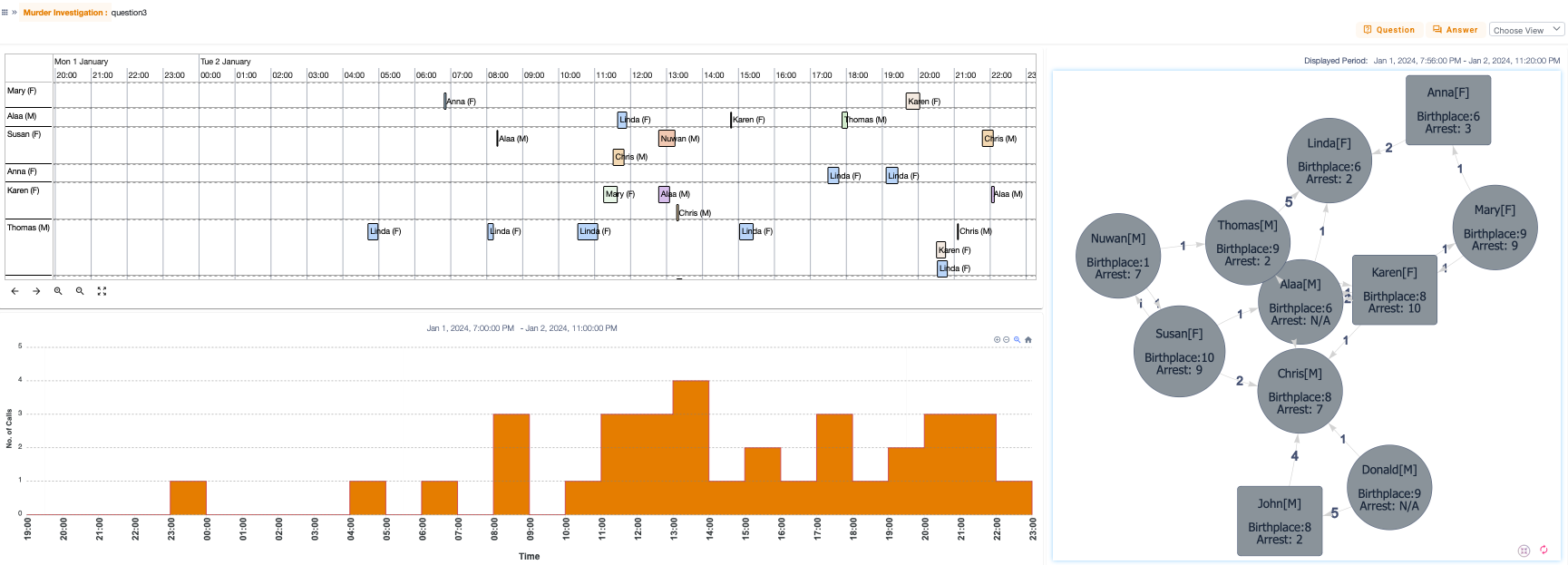}
    \caption{Crime Investigation Dashboard.}
    \label{dashboard}
    \vspace{0.5cm}
\end{figure}

The CID includes a distribution chart, a timeline, and a network graph (\Cref{dashboard}). A graphical representation of the volume of calls within the prefered time window is provided by the distribution chart. The timeline offers a breakdown of time-related metrics, focusing on the frequency and duration of calls, and the involved parties. Finally, the network graph visualization, using undirected graphs, presents information about criminal records, including arrest numbers, gender, birthplace, and incarceration states, thereby aiding in the visual identification of latent relationships and outliers.

These visualisation layouts (i.e., graph, distribution, and timeline) were selected for their ecological validity they provide (as they often appear in production-ready visualisation platforms such as the AVT), supported by a range of real-world visualization scenarios. Each visualization was developed following user-interface adaptation and information visualisation principles, systematically introducing variations to study their effect on participants' interactions with \gls{infovis} environments and subsequent task performance. Said visualization principles include well-studied visual primitives such as color, size, area, and line patterns, which map to different layout-specific adaptations to visual elements (e.g., color, shape, size, thickness, position), depending on the context and setting. Subsequently, the \gls{rl} agent is trained to select the optimal adaptation in real time.

\subsection{Data Collection}
\label{sec:data_collection}

A total of 120 participants (71 male, 49 female, aged 17–54) were recruited for a 1h study with a 25€ voucher as compensation.
The goal is to train a dedicated \gls{rl} agent for each visualization type. A dataset per type was crafted, each with data from disjoint sets of 40 participants (between-subjects design). 
The visualizations included the original versions and modifications designed to improve performance, featuring fully and partially adapted content. 
After each presentation, participants completed a question-answering task based on the displayed content. 
This allowed the final dataset to be categorized according to combinations of question and visual complexities (i.e., low vs. high), and enabled quantification of task execution efficiency in terms of accuracy and reaction time. 
Neurophysiological and behavioral signals (i.e. EEG and \gls{et}) were recorded jointly for the neurobiological definition of a \gls{mwl} proxy.


\section{System Description}
\label{sec:system}

The proposed framework employs the Event-Condition-Action paradigm to model and execute context-aware UI adaptations~\cite{Motti13} for the AEGIS' AVT dashboard (\Cref{sec:case_study}). This loop consists of a driver that connects, in real time, to the measurement devices and minimally preprocesses the raw data in 2-second epoch cycles--the Physiological Sensing (PS) module (\Cref{subsec:physiological_sensing_module}) and the Behavior Sensing  (BS) module that captured events from the dashboard such as the layout used by the user in real time. This is followed by the Compact Data Representation (CDR) module (\Cref{subsec:cdr_module}), which is responsible for engineering the cognitive features from the raw signals. Next, the Neural-Mechanistic Understanding (NMU) module (\Cref{subsec:nmu_module}) infers the \gls{mwl} state from the derived features, which is sent to the \gls{rl} module (\Cref{subsec:rl_module}), which will compute the adaptation and send it to the Adaptation Engine (AE) module (\Cref{subsec:adaptations_engine_module}) responsible for executing the changes in the dashboard.

\begin{figure}[!htbp]
    \begin{center}
    \includegraphics[width=0.99\linewidth]{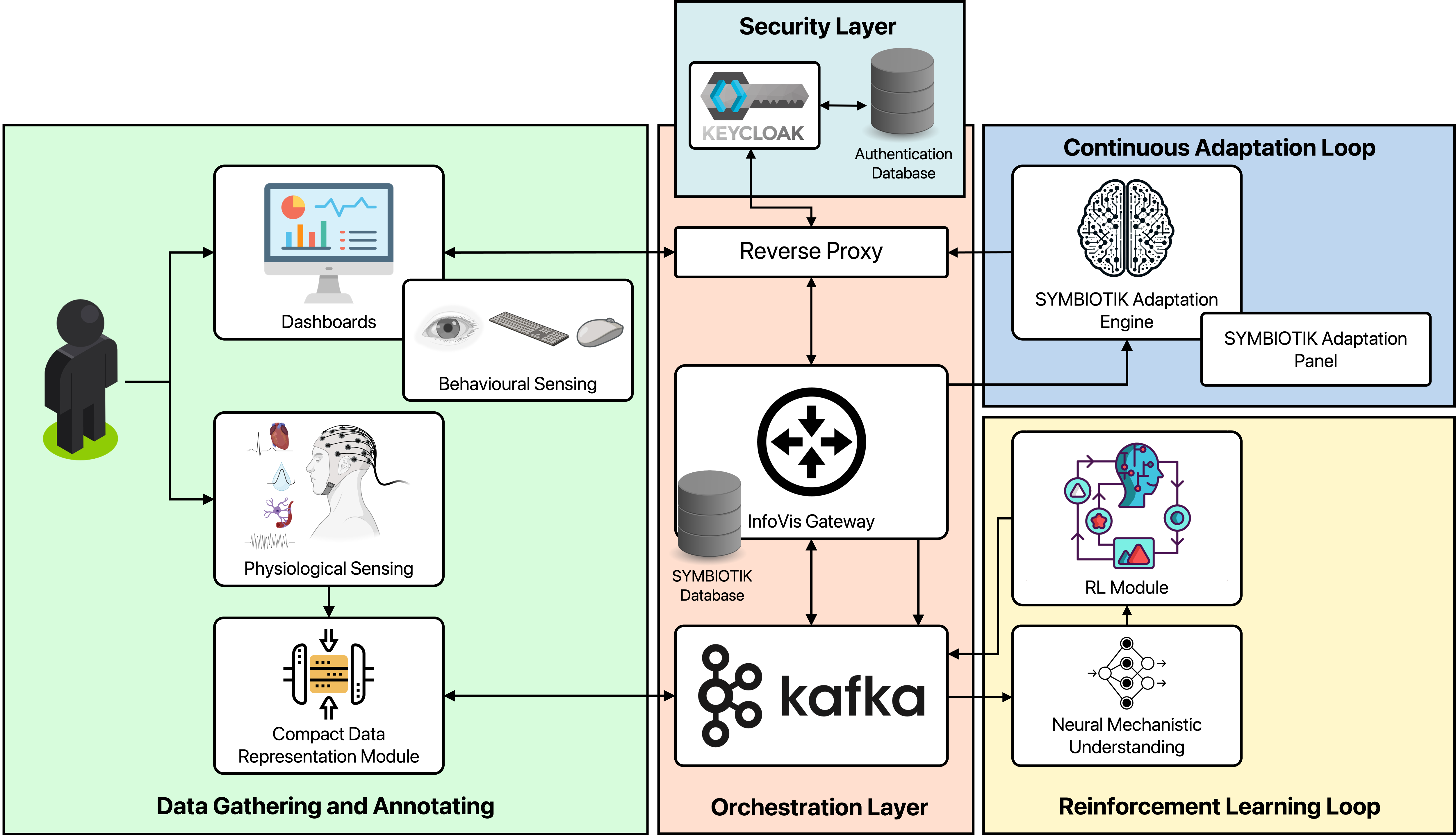}
    \end{center}

    \caption{\textsc{Symbiotik} high-level software architecture.}
    \label{fig:architecture_all}
\end{figure}




\subsection{Physiological Sensing Module}
\label{subsec:physiological_sensing_module}
This module interfaces with measurement devices, transmitting raw data in 2-second epochs. Specific Python libraries, like \textit{socket}, enable the connection to different \gls{eeg} devices and other neurophysiological signal sources, including \gls{et} that allow future integration. 

\subsection{Compact Data Representation}
\label{subsec:cdr_module}

\begin{figure}[!htbp]
    \begin{center}
       \vspace{-8pt}
    \includegraphics[width=0.8\linewidth]{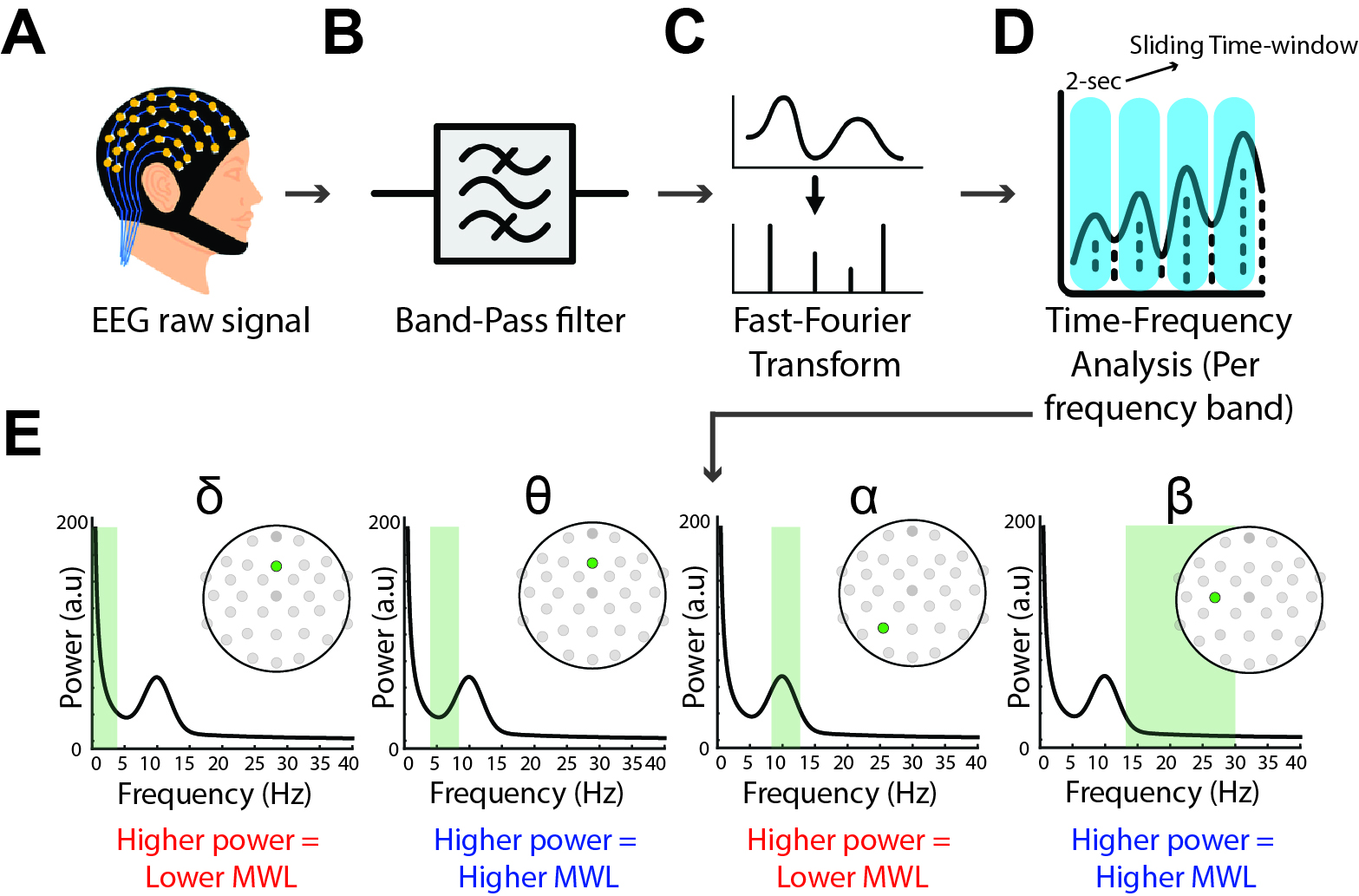}
    \end{center}
    \vspace{-8pt}
    \caption{EEG preprocessing pipeline: (A) Raw signal recording; (B) Bandpass filtering; (C) FFT to convert to frequency domain; (D) Time-frequency analysis to compare every 2 seconds the power of the following frequency bands: (E) Delta, Theta, Alpha, and Beta.}
    \label{fig:eeg_preprocessing}
\end{figure}

The raw data are then processed in the compact data representation module. Minimal preprocessing is performed to support real-time inference. Band-pass filtering (0.5--40 Hz) is performed, followed by power spectral density which is computed by taking the squared magnitude of the Fast-Fourier transformation (FFT). The resulting power is computed for the frequency bands of interest using electrodes P3 (alpha), C3 (beta), and Fz (theta and delta) (\Cref{fig:eeg_preprocessing}).

\subsection{Neural-Mechanistic Understanding Module}
\label{subsec:nmu_module}

Building on prior work \cite{kutafina2021,puma2018,so2017}, this module defines a preliminary \gls{mwl} construct by categorizing each band’s activity into low, medium, and high levels based on population-level quantiles from 120 samples. The 25th and 75th percentiles from the training data set thresholds for inference. For bands positively correlated with \gls{mwl} (theta, beta), values below the 25th percentile are considered low, above the 75th are high, and in between are medium or optimal. For negatively correlated bands (alpha, delta), this classification is reversed. Categories per band are linearly combined with fixed weights, pending further optimization.
\subsection{Reinforcement Learning Module}
\label{subsec:rl_module}
 
A \gls{rl} framework was formulated as a smart system, where an \gls{rl} agent interacts with the user and the dashboard, recommending adaptation actions to maximize cumulative rewards. Specifically, a separate agent is trained for each type of layout (graph, timeline, or distribution). To determine which agent to use at inference time, the system detects which representation the user is currently viewing through the BS module. 
The \gls{rl} module retrieves a configuration file via an API to determine the current dashboard adaptation and the question being addressed. Additionally, a dedicated classifier is trained to predict the difficulty level of the question at inference time, using popular pre-trained lightweight models for sentence classification fine-tuned on our original questions 
based on BERT \cite{devlin_bert_2019} and RoBERTa \cite{liu_roberta_2019}.
For each representation, a behavior policy is learned using the corresponding dataset (\Cref{sec:data_collection}). The reward measures whether the selected action results in an optimal \gls{mwl}. During training, an importance-weighted estimate of the reward is used, accounting for differences between the behaviour policy that generated the dataset and the target policy the agent is learning \cite{fujimoto_off-policy_2019}. The agent learns an optimal policy using a tabular Q-learning method \cite{watkins_q-learning_1992} with epsilon-greedy exploration. The algorithm operates in an environment where episodes consist of a single transition, and learning happens iteratively through a sequence of updates based on observed rewards. During training, the agent updates its Q-table (a matrix of size $n_{states} \times n_{actions}$) based on the experiences in the dataset. 
The trained agent receives the current state as input and outputs the most appropriate action. The action space (\textbf{A}) encompasses all possible adaptations within each representation. The selected action,  with its active representation, is then forwarded to the adaptation engine.

\subsection{Adaptation Engine}
\label{subsec:adaptations_engine_module}
The Adaptation Engine (AE) receives an adaptation strategy suggested by the \gls{rl} module along with a visualization type (e.g., network, timeline, distribution). Adaptation strategies are categorized as "No Adaptation", "Partial Adaptation" (i.e., with attributes such as color, shape, size, proximity, and thickness), or "Full Adaptation"~\cite{Eloi24,Todi21}. It then queries the Adaptation Catalogue, which maps a strategy to actionable adaptation operations based on ergonomic criteria ~\cite{bastien:ergonomic_prel} and linguistic levels~\cite{jean:ergo:linguistic}. The AE encodes the most appropriate operation in a JSON configuration file to be sent to the AVT dashboard via the InfoVis Gateway to adapt the final user interface accordingly, which is in turn subject to evaluation by EEG~\cite{Gaspar23}.
\subsection{Infovis Gateway}
\label{subsec:infovis_module}
This component serves as a middleware between the AP and the Dashboard. It serves as a centralized orchestrator, combining WebSocket protocols for real-time dashboard updates with Kafka~\footnote{\url{https://kafka.apache.org/}} message broker to manage high-volume data streams between the components. Kafka Broker decouples services, enabling asynchronous event handling. Each component publishes its information to Kafka topics that are organized by conceptual categories (e.g., biosignals, strategies, adaptations). Any component interested in a particular type of data can subscribe to the relevant topic, consume updates as they are published, and react autonomously.  This event-driven architecture significantly improves system scalability, flexibility, observability, real-time adaptability, and modularity. Components can operate independently without requiring knowledge of each other’s existence, facilitating easier integration of new modules and allowing the system to dynamically adapt to evolving requirements. 


\section{Conclusions and Future Work}
\label{sec:conclusions}

\textsc{Symbiotik}\footnote{\label{repo}\url{https://github.com/symbiotik-viz/prototype}} implements a scalable, real-time adaptation architecture, and a validated methodology for neuroadaptive user interfaces, by training RL agents with real-world data across different layout types. An important avenue for future work involves developing a robust and personalized \gls{mwl} metric. Preliminary analysis of our dataset revealed that behavioural indicators (e.g., accuracy and response time) show great promise in predicting question and visual complexities, as well as the effects of adaptation. These findings suggest that a multimodal approach leveraging multiple signals may help define a better \gls{mwl} proxy with real-time applicability, e.g., through strategies such as latent variable modelling. Also, using \gls{et} to indicate visual attention is a key research direction~\cite{LopezCardona2025ScanpathModels}. It reflects the perceived importance of design elements~\cite{cartella_trends_2024} and supports the development of task-specific cognitive state models generalizable across participants~\cite{langner_cognitive_2024} and layouts. The final aim is to create a common and layout-agnostic agent, able to generalize in diverse settings.



\begin{ack}
This research is supported by Horizon Europe's European Innovation Council through the Pathfinder program (SYMBIOTIK project, grant 101071147) and by the Industrial Doctorate Plan of the Department of Research and Universities of the Generalitat de Catalunya, under Grant AGAUR 2023 DI060.
\end{ack}



\bibliography{bib/infovis_paper, bib/rl, bib/neural_eeg, bib/AE}


\end{document}